\documentclass[10pt,a4paper,twoside,twocolumn,english,pra,onecolum]{revtex4-1}
\usepackage[T1]{fontenc}
\usepackage[latin9]{inputenc}
\usepackage{amsmath}
\usepackage{amsthm}
\usepackage{amssymb}
\usepackage[pdftex]{graphicx}

\makeatletter

\pdfpageheight\paperheight
\pdfpagewidth\paperwidth

\newcommand{\lyxdot}{.}

\makeatother

\usepackage{babel}
\begin{document}

\title{Quantum sensing of rotation velocity based on transverse field Ising
model}

\author{Y. H. Ma$^{1,2}$ C. P. Sun$^{1,2}$}
\email{cpsun@csrc.ac.cn}

\address{$^{1}$Beijing Computational Science Research Center, Beijing 100193,
China~\\
 $^{2}$Graduate School of Chinese Academy of Engineering Physics,
Beijing 100084, China}
\begin{abstract}
We study a transverse-field Ising model (TFIM) in a rotational reference
frame. We find that the effective Hamiltonian of the TFIM of this
system depends on the system's rotation velocity. Since the rotation
contributes an additional transverse field, the dynamics of TFIM sensitively
responses to the rotation velocity at the critical point of quantum
phase transition. This observation means that the TFIM can be used
for quantum sensing of rotation velocity that can sensitively detect
rotation velocity of the total system at the critical point. It is
found that the resolution of the quantum sensing scheme we proposed
is characterized by the half-width of Loschmidt echo of the dynamics
of TFIM when it couples to a quantum system S. And the resolution
of this quantum sensing scheme is proportional to the coupling strength
$\delta$ between the quantum system S and the TFIM , and to the square
root of the number of spins $N$ belonging the TFIM.
\end{abstract}
\maketitle

\section{Introduction}


Quantum sensing is a kind of sensing scheme, which utilizes the quantum
effects to enhance the measurement accuracy. Quantum gyroscope \cite{key-1}
is a kind of quantum sensor which makes use of quantum sensing schemes
for rotation velocity measurement. For example, atom interferometer
gyroscope (AIG) makes use of the interference of matter waves, and
was first achieved in experiment in 1991\cite{key-2}. Another quantum
gyroscope is nuclear magnetic resonance gyroscope (NMRG), which detects
the rotation velocity by detecting the precession frequency of the
nuclear magnetic moment in the non-inertial system\cite{key-3}. Both
AIG and NMRG are quantum gyroscopes with high accuracy in measurement,
and are used in high precision inertial navigation and military strategic
system.

For a future quantum sensing scheme with much higher accuracy, we
need to explore the roles which could be positive or negative of new
quantum effects. The key to the quantum sensing scheme is the measurement
accuracy, or in our case, the resolution for rotation velocity, which
is physically reflected by the response of its dynamics to the rotation
velocity of the system. In this paper, we study the dynamics of transverse
field Ising model (TFIM) and found the Loschmidt echo (LE) of TFIM
is sensitive to the rotation velocity at the critical point of quantum
phase transition (QPT). In 2006, H.T.Quan, et al \cite{key-4} found
that when a TFIM couples with a quantum system S, and is turned into
critical point of QPT, the quantum decoherence phenomena of system
S will be significantly enhanced, which is characterized by the rapid
decay of LE of the dynamics of TFIM around the critical point. For
TFIM in non-inertial reference frame, its quantum phase transition
behavior depends on the rotation velocity $\varOmega$ of the total
system sensitively as the $\varOmega$ behaves as an external transverse
field.

The QPT of TFIM in rotational reference frame will be affected directly
by the rotation velocity of the system, therefore it is feasible to
achieve the measurement of rotation velocity through recording the
QPT of TFIM. With this consideration, we designed a quantum sensing
scheme to carry out rotation velocity measurements by detecting the
QPT of TFIM in non-inertial reference frame. This paper is arranged
as follows. In Sec. II, we study the Hamiltonian of the TFIM in non-inertial
system and thus examine the response of the TFIM\textquoteright s
LE to the rotation velocity $\varOmega$ of the system. In Sec. III,
we propose a quantum sensing scheme based on TFIM in rotational reference
frame, and give its resolution for rotation velocity $\varOmega$.
Conclusion of our results are given in Sec. IV.


\section{transverse field ising model in non-inertial reference frame}



In this section we consider a coupled spin system with an external
field in a rotational reference frame. The Hamiltonian of the transverse
field Ising model (TFIM) in the stationary system reads

\begin{equation}
H_{0}=-J\sum_{i}\text{\ensuremath{\left(\sigma_{i}^{z}\sigma_{i+1}^{z}+\lambda\sigma_{i}^{x}\right)}},\label{eq:H0}
\end{equation}
where $J$ and $\lambda$ characterize the strengths of the inter-spin
interaction and the coupling to the transverse field, $\sigma_{i}^{\alpha}\left(\alpha\in\left\{ x,y,z\right\} \right)$are
Pauli operators defined in the state space of the $i$th particle.
Note that, the critical point for quantum phase transition (QFT) of
TFIM is $\lambda_{c}=1$ \cite{key-4,key-5}, below which the TFIM
will generate spontaneous magnetization . While a spin chain of TFIM
is placed in a rotating system which rotates with an angle $\theta\left(t\right)$
relative to the stationary reference frame, the time-dependent Hamiltonian
of TFIM is

\begin{equation}
H\left(t\right)=-J\sum_{i}\text{\ensuremath{\left(\sigma_{i}^{z}\left(t\right)\sigma_{i+1}^{z}\left(t\right)+\lambda\sigma_{i}^{x}\left(t\right)\right)}},\label{eq:hamiltoniant}
\end{equation}
where $\sigma_{i}^{\alpha}\left(t\right)$ are Pauli matrices in the
coordinate system which is relatively stationary with the TFIM. With
the help of rotation matrix $D_{\overrightarrow{n}}\left(\theta\right)$,
$\sigma_{i}^{\alpha}\left(t\right)$ can be explicitly expressed by
$\sigma_{i}^{\alpha}$

\begin{equation}
\left(\begin{array}{c}
\sigma_{i}^{x}\left(t\right)\\
\sigma_{i}^{y}\left(t\right)\\
\sigma_{i}^{z}\left(t\right)
\end{array}\right)=D_{\overrightarrow{n}}\left(\theta\right)\left(\begin{array}{c}
\sigma_{i}^{x}\\
\sigma_{i}^{y}\\
\sigma_{i}^{z}
\end{array}\right),\label{eq:rotationoperation}
\end{equation}
where $\overrightarrow{n}$ is the direction vector of the rotation
axis. When the system rotates counterclockwise in the x-direction,
the rotation matrix is

\begin{equation}
D_{x}\left(\theta\right)=\left(\begin{array}{ccc}
1 & 0 & 0\\
0 & \cos\theta & -\sin\theta\\
0 & \sin\theta & \cos\theta
\end{array}\right).\label{rotationmatrix}
\end{equation}
In addition, we set $\theta\left(0\right)=0$, so that $H\left(0\right)=H_{0}$.

\begin{center}
\begin{figure}
\includegraphics[scale=0.2]{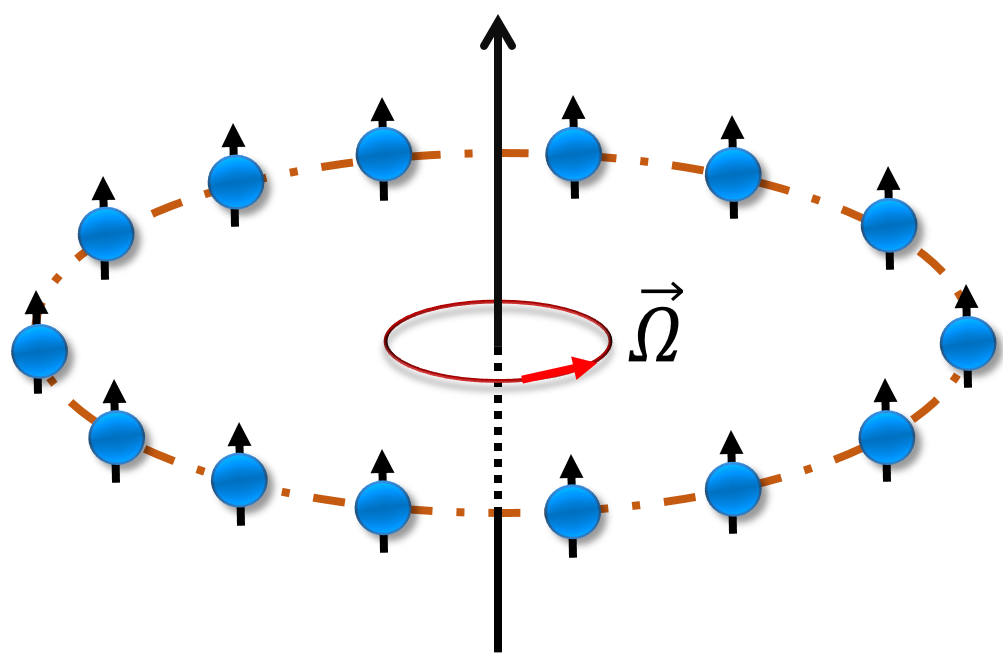}

\raggedright{}FIG. 1 (color online). Transverse field Ising model
in rotation reference frame. The whole system is rotating around the
axis with rotation velocity $\overrightarrow{\varOmega}$.
\end{figure}
\par\end{center}

The Schrodinger's equation in the rest reference frame for TFIM with
rotation angle $\theta\left(t\right)$ is $i\hbar\partial\varPsi/\left(\partial t\right)=H\left(\theta\right)\varPsi$,
where $\varPsi$ is the wave function of TFIM in the rest reference
frame. Obviously, $\varPsi^{'}=R\left(\theta\right)\varPsi$ represents
the wave function of TFIM in the rotation reference frame, where $R\left(\theta\right)=\exp\left(-i\overrightarrow{\theta}\cdot\overrightarrow{S}/2\right)$
is the rotation operator and $\overrightarrow{S}=\sum_{i}\overrightarrow{s_{i}}$
is the total spin of the TFIM, while $\overrightarrow{s_{i}}$ is
the spin of the $i$th particle. Thus the Schrodinger's equation for
$\varPsi^{'}$is

\begin{equation}
i\hbar\frac{\partial}{\partial t}\varPsi^{'}=\left(R\left(\theta\right)H\left(t\right)R^{\dagger}\left(\theta\right)-i\hbar R\left(\theta\right)\frac{\partial}{\partial t}R^{\dagger}\left(\theta\right)\right)\varPsi^{'}.\label{eq:se}
\end{equation}
Eq. (\ref{eq:se}) indicates the effective Hamiltonian of the rotating
TFIM in non-inertial reference is

\begin{equation}
H_{eff}=R\left(\theta\right)H\left(t\right)R^{\dagger}\left(\theta\right)-i\hbar R\left(\theta\right)\frac{\partial}{\partial t}R^{\dagger}\left(\theta\right).\label{eq:Heff}
\end{equation}
For the system rotating around x axis, $R\left(\theta\right)$ is

\begin{equation}
R_{x}\left(\theta\right)=\exp\left(-i\theta\sum_{i}\sigma_{i}^{x}/2\right).\label{eq:ro}
\end{equation}
It follows from Eqs. (\ref{eq:hamiltoniant}), (\ref{eq:rotationoperation}),
(\ref{rotationmatrix}), (\ref{eq:Heff})and (\ref{eq:ro}) that the
effective Hamiltonian for the rotating TFIM in non-inertial reference
frame is obtained as

\begin{equation}
H_{eff}=-J\sum_{i}\text{\ensuremath{\left[\sigma_{i}^{z}\sigma_{i+1}^{z}+\left(\lambda-\frac{\hbar\varOmega}{2J}\right)\sigma_{i}^{x}\right]}}.\label{eq:Heff1}
\end{equation}
Here $\varOmega=d\theta/\left(dt\right)$ indicates the instantaneous
rotation velocity of the total system. It is imagined from Eq. (\ref{eq:H0})
that the rotation term in Eq. (\ref{eq:Heff1}) can be regarded as
an effective magnetic field interacting with spins in the x direction.
For a single spin in TFIM, if we ignore the interaction with other
spins, the effective Hamiltonian will be $H_{eff}^{i}=-J\left(\lambda-\frac{\hbar\varOmega}{2J}\right)\sigma_{i}^{x}$,
which is the same as that of the nuclear magnetic resonance(NMR).
The physical meaning of this rotation term is similar to the Coriolis
force, which is known as a basic non-inertial effect in classical
mechanics.

In 2006, H.T.Quan et al. pointed out that \cite{key-4}, when TFIM
couples with a two level quantum system S, the decoherence of S will
be effectively enhanced while the TFIM is at its critical point for
QPT. This phenomenon was proved to be directly related to the Loschmidt
echo (LE) of the dynamical behavior of TFIM and has been observed
experimentally\cite{key-6,key-7}. The Hamiltonian they used to describe
the interaction between S and TFIM is

\begin{equation}
H\left(\lambda,\delta\right)=-J\sum_{i}\text{\ensuremath{\left(\sigma_{i}^{z}\sigma_{i+1}^{z}+\lambda\sigma_{i}^{x}+\delta\left|e\right\rangle \left\langle e\right|\sigma_{i}^{x}\right)}},\label{eq:hamiltonian}
\end{equation}
where $J$,$\lambda$,$\sigma_{i}^{\alpha}\left(\alpha\in\left\{ x,y,z\right\} \right)$
are consistent with our previous explanation, and $\delta$ is the
coupling strength between TFIM and quantum system S. If the whole
system is placed in a rotating system, there will be a correction
to Hamiltonian in Eq. (9), which is

\begin{equation}
H\left(\widetilde{\lambda},\delta\right)=-J\sum_{i}\text{\ensuremath{\left(\sigma_{i}^{z}\sigma_{i+1}^{z}+\widetilde{\lambda}\sigma_{i}^{x}+\delta\left|e\right\rangle \left\langle e\right|\sigma_{i}^{x}\right)}}.\label{eq:hamiltoniane}
\end{equation}
Here,

\begin{equation}
\widetilde{\lambda}\equiv\lambda-\frac{\hbar\varOmega}{2J},\label{eq:lambdae}
\end{equation}
is defined as the total effective magnetic field which is modified
by the rotation velocity $\varOmega$ of the non-inertial reference
frame. To show the effect of the rotation on the dynamic of the system,
we calculate the LE of TFIM \cite{key-4} by taking $N=2000$, $\delta=0.01$,
$\lambda=2$, $\hbar=1$, $\varOmega$$\in[0,4J]$. The result we
get is demonstrated in FIG.2, which shows the relationship between
LE and the effective magnetic field $\widetilde{\lambda}$.

\begin{center}
\begin{figure}
\includegraphics[scale=0.4]{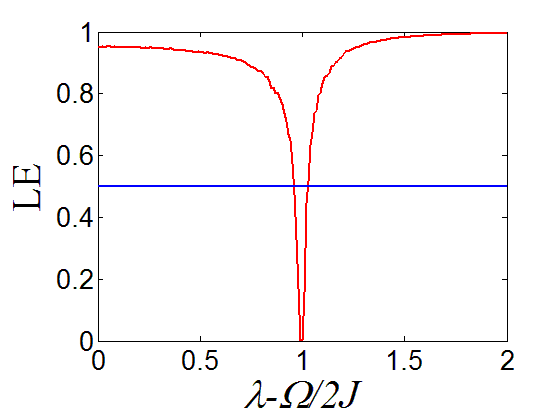}
\raggedright{}FIG. 2 (color online). Loschmidt echo as a function
of rotation velocity $\varOmega$ for system with $N=2000$, $\delta=0.01$.
This figure shows that around the critical point $\widetilde{\lambda}=1$,
a tiny change of $\varOmega$ will cause rapid change of LE. The blue
line in the figure marks $L=0.5$ , and the length of which inside
the valley of LE represent the half-width of LE. The half-width of
Loschmidt echo demonstrates the resolution of LE to the change of
rotation velocity and will be further discussed in Section III.B
\end{figure}
\par\end{center}

It is obvious in FIG.2 that a slight change of rotation velocity around
the critical point for QFT of TFIM($\widetilde{\lambda}=\lambda-\hbar\Omega/(2J)=1$)
results in a significant change in the value of LE, which indicates
that the dynamics of of TFIM is sensitive to the rotation velocity
at the critical point of QPT.This suggests that TFIM can serve as
a gyroscope in principle to measure the rotation velocity of a non-inertial
system.

\section{Quantum sensing scheme for rotation velocity }

Due to the sensitive dependence of the parameters($\lambda,\varOmega$)
to QPT of TFIM, there comes an idea that the rotation of a system
can be detected through the observation of TFIM's QPT. In another
word, we propose a quantum sensing scheme which makes use of the quantum
phase transition effect of TFIM. It can be seen in FIG. 2 that the
significant response of LE to the change of rotation velocity occurs
only near the critical point for QPT of TFIM.

To utilize the QPT effect to detect the rotation velocity sensitively,
we need to keep $\widetilde{\lambda}$ near the critical point $\lambda_{c}=1$.
But for any rotation velocity to be measured, $\widetilde{\lambda}$
may be far away from $\lambda_{c}$, thus we need to adjust the magnetic
field $\lambda$ to ensure that $\widetilde{\lambda}$ fall in the
range we are expecting. For the sensing scheme to be working properly,
we need the magnetic field to be adjusted into the vicinity area of
$\widetilde{\lambda}$, but the problem is we do not know the value
of $\widetilde{\lambda}$ in advance. To solve this problem, we present
the following measurement scheme, which is illustrated in Fig. 3.
\begin{figure}
\begin{centering}
\includegraphics[scale=0.4]{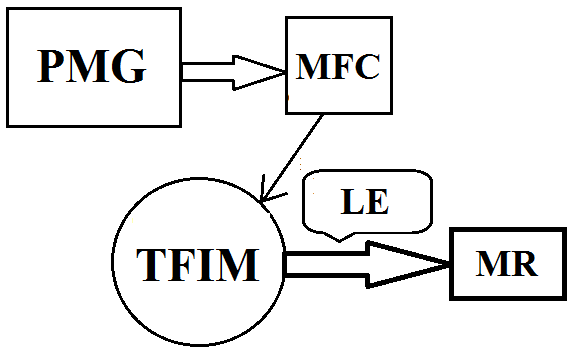}
\par\end{centering}
\raggedright{}FIG. 3. Workflow of the quantum sensing scheme. The
entire quantum sensing scheme needs to be achieved by the following
modules. A pre-measuring gyroscope (PMR), which may be a MEMS gyroscope
or a fiber optic gyroscope or other kinds of classical gyroscopes,
a magnetic filed controller (MFC), a transverse field Ising model
(TFIM) and circuitry to connect the entire system.
\end{figure}

First, we use a pre-measuring gyroscope (PMG) to carry out a pre-measurement
and send the result $\varOmega_{0}$ to the magnetic filed controller
(MFC). Then the MFC will adjust the amplitude of the magnetic filed
inside the module of TFIM to make the effective magnetic field $\widetilde{\lambda}$
in the vicinity of the critical point that $\widetilde{\lambda}\approx\lambda_{c}=1$.
Let $\sigma$ be the resolution of PMR, then the rotation velocity
to be measured is in the following range

\begin{equation}
\varOmega_{1}\in\left[\varOmega_{0}-\sigma,\varOmega_{0}+\sigma\right].\label{eq:range}
\end{equation}
From Eqs. (\ref{eq:lambdae}), (\ref{eq:range}), we obtain the magnetic
field adjustment range of MFC

\begin{equation}
1+\frac{\hbar}{2J}\left(\varOmega_{0}-\sigma\right)\leq\lambda\leq1+\frac{\hbar}{2J}\left(\varOmega_{0}+\sigma\right),
\end{equation}
which is the key to make $\widetilde{\lambda}$ change in the vicinity
of TFIM's critical point. When the QPT of TFIM occurs at $\widetilde{\lambda}=\lambda_{c}=1$,
the LE will rapidly decay to zero. At the same time, the amplitude
of the magnetic field is denoted as $\lambda=\lambda_{0}$. Thus the
rotation velocity is measured as a result that

\begin{equation}
\varOmega_{1}=\frac{2J}{\hbar}\left(\lambda_{0}-1\right).
\end{equation}

However, if we want to use this quantum sensing scheme to achieve
a meaningful measurement for rotation velocity(the resolution is improved
by TFIM compared with that of PMG, the distinguish-ability of TFIM's
quantum phase transition to rotation velocity should be higher than
that of PMR, which is to say

\begin{equation}
\triangle\varOmega<\sigma,\label{eq:condition}
\end{equation}
where $\triangle\varOmega$ is the resolution of TFIM's QPT to rotation
velocity.

For a quantum sensing scheme, its resolution is an important parameter,
which indicates the minimum rotation velocity it can measure. In the
above scheme we have proposed, the resolution of TFIM's QPT to rotation
velocity is characterized by the half-width of LE of the dynamics
of TFIM. The smaller the LE's half-width is, the smaller rotation
velocity the system can discern, and thus the higher resolution the
system pocesses. The exact expression for LE of the system with Hamiltonian
in Eq. (\ref{eq:hamiltonian}) was given in paper \cite{key-4}. It
can be seen from Eq. (\ref{eq:hamiltoniane}) that, while the entire
system is placed in non-inertial reference frame, the Hamiltonian
has the same form as in Eq. (\ref{eq:hamiltonian}), except $\lambda$
is replaced by $\widetilde{\lambda}$. By using the constant variable
method, we make the replacement $\lambda\rightarrow\widetilde{\lambda}$
for the solutions given in paper {[}4{]}. Thus the exact expression
for $L\left(\widetilde{\lambda,}t\right)$ is naturally obtained as

\begin{equation}
L\left(\widetilde{\lambda,}t\right)=\prod_{k>0}[1-\sin^{2}(2\alpha_{\widetilde{\lambda},k})\sin^{2}(\varepsilon_{e}^{\widetilde{\lambda},k}t)],\label{eq:L}
\end{equation}
where

\[
\alpha_{\widetilde{\lambda},k}=\frac{1}{2}\left[\theta_{\widetilde{\lambda},k}\left(0\right)-\theta_{\widetilde{\lambda},k}\left(\delta\right)\right],
\]
and

\[
\theta_{\widetilde{\lambda},k}\left(\delta\right)=\arctan\left\{ -\sin\left(ka\right)/\left[\cos(ka)-\left(\widetilde{\lambda}+\delta\right)\right]\right\} ,
\]
and the single quasiexciation energy

\[
\varepsilon_{e}^{\widetilde{\lambda},k}\left(\delta\right)=2J\sqrt{1+\left(\widetilde{\lambda}+\delta\right)^{2}-2\left(\widetilde{\lambda}+\delta\right)\cos\left(ka\right)}.
\]
Here, the Bloch wave vector $k$ takes the discrete values $2n\pi/\left(Na\right)\left(n=1,2,\ldots,N/2\right)$,
where $a$ and $N$ are the lattice spacing and particle number of
TFIM.We first make an analytical analysis by considering the partial
sum with a cutoff wave vector $K_{c}$, thus

\begin{equation}
S\left(\widetilde{\lambda},t\right)=\ln L_{c}=-\sum_{k>0}^{K_{c}}\left|\ln F_{k}\left(\widetilde{\lambda},t\right)\right|,\label{eq:S}
\end{equation}
where $F_{k}\left(\widetilde{\lambda},t\right)=1-\sin^{2}(2\alpha_{\widetilde{\lambda},k})\sin^{2}(\varepsilon_{e}^{\widetilde{\lambda},k}t)$,
and $K_{c}$ can be expressed by a cutoff number $N_{c}$ as $K_{c}=N_{c}\pi/\left(Na\right)$.
When $K_{c}a\ll1\left(N_{c}\ll N\right)$

\begin{equation}
S=-\frac{\delta^{2}m\sin^{2}\left[2J\left(1-\widetilde{\lambda}\right)t/\hbar\right]}{\left(1-\widetilde{\lambda}\right)^{2}\left(1-\widetilde{\lambda}-\delta\right)^{2}},\label{eq:S1}
\end{equation}
where

\begin{equation}
m\equiv\frac{4\pi^{2}N_{c}\left(N_{c}+1\right)\left(2N_{c}+1\right)}{6N^{2}}.\label{eq:m}
\end{equation}
It follows from Eqs. (\ref{eq:S}), (\ref{eq:S1}) that

\begin{equation}
L_{c}=\exp\left\{ -\frac{\delta^{2}m\sin^{2}\left[2J\left(1-\widetilde{\lambda}\right)t/\hbar\right]}{\left(1-\widetilde{\lambda}\right)^{2}\left(1-\widetilde{\lambda}-\delta\right)^{2}}\right\} .
\end{equation}
Around the critical point for QPT, we set $\epsilon=1-\widetilde{\lambda}-\delta$,
as a result, if $2J\left(\epsilon+\delta\right)t/\hbar\ll1$ we have

\begin{flushleft}
\begin{equation}
L_{c}\left(\epsilon,t\right)=e^{-4\frac{\delta^{2}}{\epsilon^{2}}mJ^{2}t^{2}/\hbar^{2}}.
\end{equation}
To get the time-independent half-width of $L_{c}$, we define the
characteristic time
\par\end{flushleft}

\begin{equation}
t_{0}\equiv\frac{\hbar}{2J},
\end{equation}
 then we can calculate the half-width of $L_{c}$, which is defined
as $\epsilon_{0}$, at $t=t_{0}$

\begin{flushleft}
\begin{equation}
\frac{1}{2}=L_{c}\left(\epsilon_{0},t_{0}\right)=e^{-\delta^{2}m/\epsilon_{0}^{2}}.
\end{equation}
 In this case, the half-width of $L_{c}$ is obtained as

\par\end{flushleft}\begin{center}
\begin{equation}
\epsilon_{0}=\delta\sqrt{m}/\sqrt{\ln2}\approx\delta\sqrt{m}.\label{eq:eps}
\end{equation}
\par\end{center}

For the above analytical calculation, we further assume that the momentum
cutoff $K_{c}$ is an $N$-independent constant in the limit $N\rightarrow\infty$,
i.e., we have

\begin{equation}
N_{c}=\alpha N,\label{eq:NC}
\end{equation}
where $\alpha=K_{c}a/\pi$ is also an $N$-independent constant. Later
we will verify this assumption and determine the value of $\alpha$
via exact numerical calculation. Using relation (\ref{eq:NC}), we
find that the parameter $m$ defined in Eq. (\ref{eq:m}) is expressed
as

\begin{equation}
m=\frac{4\pi^{2}\alpha N\left(\alpha N+1\right)\left(2\alpha N+1\right)}{6N^{2}}\approx\frac{4}{3}\pi^{2}\alpha^{3}N\equiv\eta N.\label{eq:mnew}
\end{equation}
Thus, the Eq. (\ref{eq:mnew}) of the half-width $\epsilon_{0}$ of
$L_{c}$ becomes

\begin{equation}
\epsilon_{0}\approx\delta\sqrt{m}=\sqrt{\eta}\delta N^{\frac{1}{2}}.\label{eq:eps0}
\end{equation}
Notice that $\eta=4\pi^{2}\alpha^{3}/3$ is an $N$-independent parameter.

Equation. (\ref{eq:eps0}) implies that the half-width $\epsilon_{0}$
or the behavior of $L_{c}$ as a function of $\tilde{\lambda}$ is
determined by $\delta N^{\frac{1}{2}}$. This conclusion can be justified
by numerical calculation by taking $\lambda=2$, $\hbar=1$, $\varOmega$$\in[0,4J]$,
and the results are shown in Fig. 4
\begin{figure}
\includegraphics[scale=0.22]{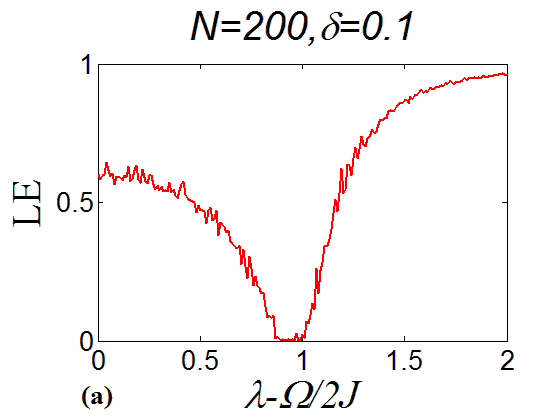}\includegraphics[scale=0.22]{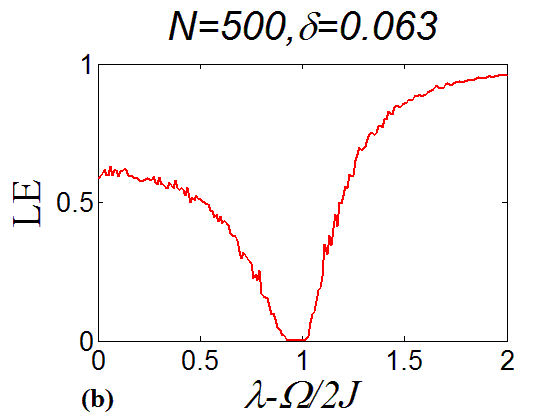}

\includegraphics[scale=0.22]{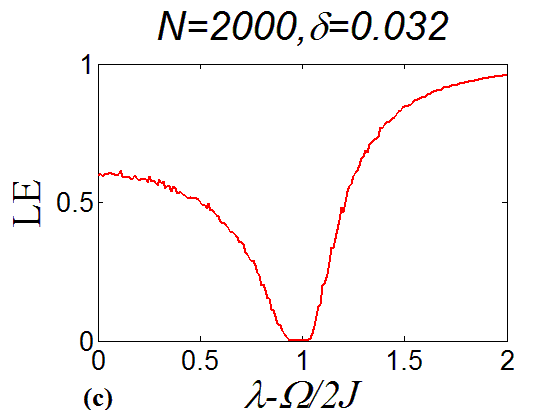}\includegraphics[scale=0.22]{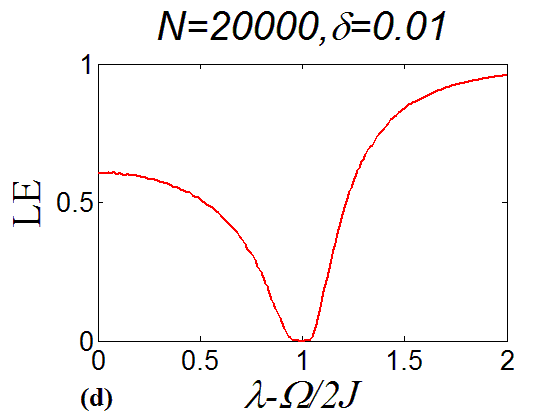}
\raggedright{}FIG. 4. (color online) Diagrams of Loschmidt echo as
a function of rotation velocity $\varOmega$ in the case where $\sqrt{N}\delta=\sqrt{2}$.
(a) System with $N=200$ and $\delta=0.1$. (b) System with $N=500$
and $\delta=0.063$. (c) System with $N=2000$ and $\delta=0.032$.
(d) System with $N=20000$ and $\delta=0.01$. These curves of LE
have the similar configuration.
\end{figure}
, in which we illustrate $L_{c}$ for the cases where the parameter
$\delta N^{\frac{1}{2}}$ is fixed at $\sqrt{2}$ while $N$ takes
different values at $200$, $500$, $2000$, and $20000$. It is clearly
shown that in these cases the behavior of $L_{c}$ are quite similar.
In Fig. 5
\begin{figure}
\includegraphics[scale=0.22]{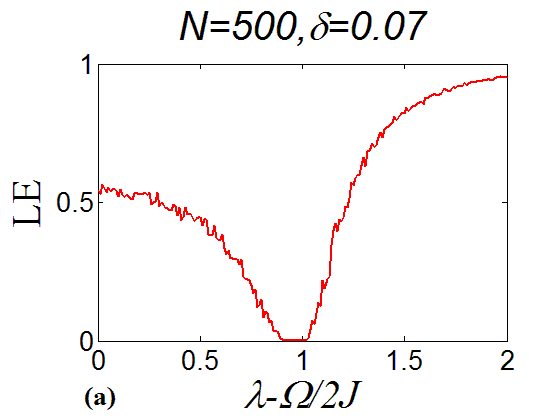}\includegraphics[scale=0.22]{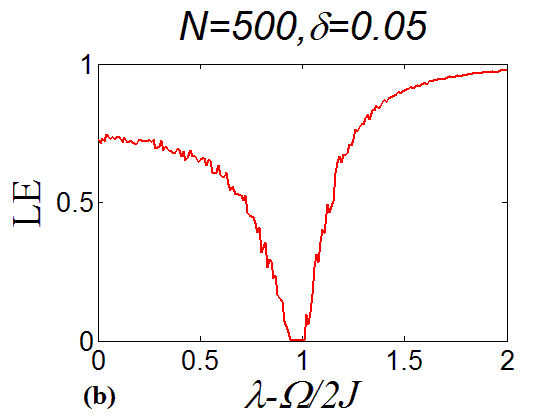}

\includegraphics[scale=0.22]{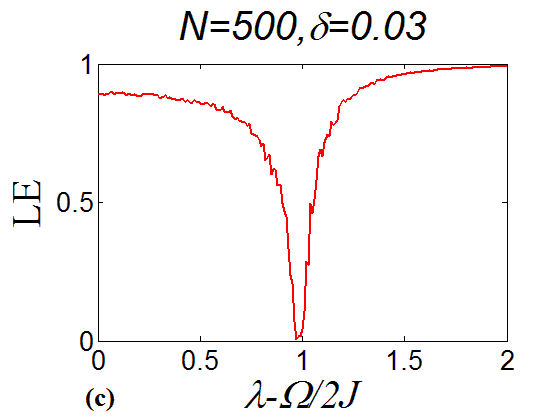}\includegraphics[scale=0.22]{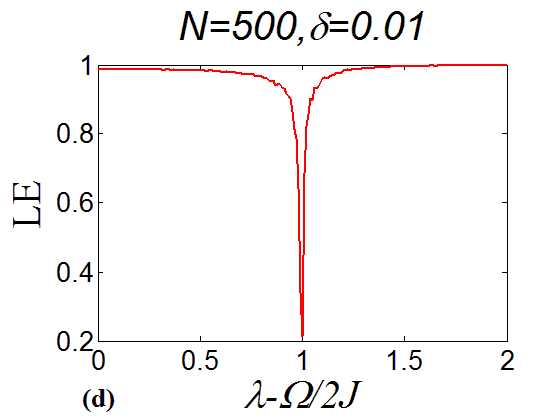}
\raggedright{}FIG. 5. (color online) Diagrams of Loschmidt echo as
a function of rotation velocity $\varOmega$ for system with $N=500$,
and $\delta$ respectively take $0.07,0.05,0.03,0.01$ in figure (a),
(b), (c) and (d). These diagrams show that the valley of the curve
become narrower when $\delta$ is decreasing
\end{figure}
we show $L_{c}$ for the cases with fixed value of $N$ and different
values of $\delta$. The calculation shows that the width of $L_{c}$
decrease significantly with $\delta$. The numerical results in the
two figures clearly confirm the results given by our analytical analysis,
i.e., the behavior of $L_{c}$ is determined by the parameter $\delta N^{\frac{1}{2}}$in
the large-$N$ limit. Furthermore, with the help of Eq. (\ref{eq:eps0})
and the exact numerical solution of $\epsilon_{0}$ given by Eq. (\ref{eq:L}),
we fit the value of parameter $\alpha$ or $\eta$ for system with
$N=20000$. As shown in Fig. 6
\begin{figure}
\includegraphics[scale=0.25]{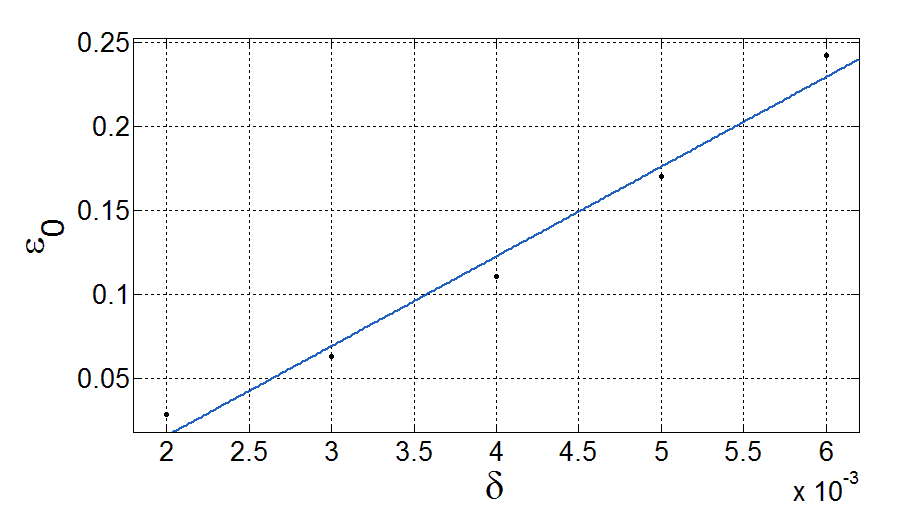}
\raggedright{}FIG. 6 (color online). Half-width of Loschmidt echo
changes with different $\delta$. The black points in this figure
is LE's half-width $\epsilon_{0}$ given by numerical calculation,
and the blue line is the fitting line of these points.It is obviously
there exist a liner relationship between $\epsilon_{0}$ and $\delta$,
and the proportionality coefficient is given as $\sqrt{\eta}=0.375$.
\end{figure}
, the fitting gives $\sqrt{\eta}=0.375$. According to Eqs. (\ref{eq:lambdae}),
(\ref{eq:eps0}), the resolution of IGS is obtained as

\begin{equation}
\triangle\varOmega=\frac{2J\epsilon_{0}}{\hbar}=2\sqrt{\eta}\omega_{0}\delta\sqrt{N}=0.75\omega_{0}\delta\sqrt{N},
\end{equation}
where $\omega_{0}\equiv J/\hbar$ is the characteristic coupling frequency
of the spins' interaction. For $\omega_{0}\sim1$Hz, $\delta=1\times10^{-5}$,
$N=2000$, $\triangle\varOmega\approx3.4\times10^{-4}\left(\textdegree\right)/s$,
accordingly, the constraint condition (\ref{eq:condition}) becomes

\begin{equation}
\delta\sqrt{N}<1.33\frac{\sigma}{\omega_{0}}.
\end{equation}
Here, $\sigma$ is the resolution of the pre-measuring gyroscope.

\section{Conclusion}

In summary, we have studied the rotation effect of the reference frame
on the quantum phase transition (QPT) of the transverse field Ising
model (TFIM). Since the rotation velocity will apply an equivalent
magnetic field to the original transverse field, the dynamic evolution
of the TFIM is sensitive to the rotation velocity of the reference
frame at the critical point of TFIM; when we adjust the original transverse
field to the vicinity of the critical point for QPT of TFIM, the Loschmidt
echo will change significantly due to small changes of rotation velocity.
This finding inspire us to design a quantum sensing scheme for measuring
rotation velocity.

The quantum sensing scheme presented in this paper is composed of
three steps. First, the approximate range of the rotation velocity
is obtained by the pre-measurement. Then the magnetic field is adjusted
based on the result of pre-measurement to tune the TFIM near the critical
point of QPT. Finally, the rotation velocity of the system is obtained
by analyzing the feature of LE. Furthermore, we found the resolution
of this quantum sensing scheme is proportional to $\delta\sqrt{N}$,
where $\delta$ is the coupling strength between quantum system S
and TFIM, and $N$ is the number of spins belongs to TFIM.
\begin{acknowledgments}
Y. H. Ma would like to thank Jin-fu Chen, Yi-nan Fang, Guo-hui Dong
and Xin Wang in Beijing Computational Science Research Center for
helpful discussion. This study is supported by the National Basic
Research Program of China (Grant No. 2014CB921403 \& No. 2016YFA0301201),
the NSFC (Grant No. 11421063 \& No. 11534002), and the NSAF (Grant
No. U1530401).
\end{acknowledgments}

\end{document}